\documentclass[useAMS,usenatbib,onecolumn]{mn2e}

\title[Fast Power]{Fast Power Spectrum Estimation}
\author{Ue-Li Pen}
\author[Ue-Li Pen]{
Ue-Li Pen,$^{1}$\thanks{E-mail:\ pen@cita.utoronto.ca}
\\
$^1$\,Canadian Institute for Theoretical Astrophysics, University of
Toronto, M5S 3H8, Canada \\}

\begin{document}
\newcommand{\etal}{{\it et al. }}
\newcommand{\beq}{\begin{equation}}
\newcommand{\eeq}{\end{equation}}
\newcommand{\Tr}{{\rm Tr}}
\date{version 11 April 2003}

\pagerange{\pageref{firstpage}--\pageref{lastpage}}
\pubyear{2003}

\newcommand{\araa}{ARA\&A}   
\newcommand{\afz}{Afz}       
\newcommand{\aj}{AJ}         
\newcommand{\azh}{AZh}       
\newcommand{\aaa}{A\&A}      
\newcommand{\aas}{A\&AS}     
\newcommand{\aar}{A\&AR}     
\newcommand{\apj}{ApJ}       
\newcommand{\apjs}{ApJS}     
\newcommand{\apjl}{ApJ}      
\newcommand{\apss}{Ap\&SS}   
\newcommand{\baas}{BAAS}     
\newcommand{\jaa}{JA\&A}     
\newcommand{\mnras}{MNRAS}   
\newcommand{\nat}{Nat}       
\newcommand{\pasj}{PASJ}     
\newcommand{\pasp}{PASP}     
\newcommand{\paspc}{PASPC}   
\newcommand{\qjras}{QJRAS}   
\newcommand{\sci}{Sci}       
\newcommand{\sova}{SvA}      
\newcommand{\rasav}{Ric.~astr.~Specola astr.~Vatic.}
\newcommand{\sca}{Scient.~Am.}   
\newcommand{\stel}{Sky Telesc.}  
\newcommand{\spsrev}{Space Sci.~Rev.} 
\newcommand{\phfl}{Phys. Fluids}
\newcommand{\phrev}{Phys. Rev.}
\newcommand{\rprph}{Rep. Prog. Phys.}
\newcommand{\rmph}{Rev. Mod. Phys.}
\newcommand{\jplph}{J. Plasma Phys.}
\newcommand{\jmph}{J. Math. Phys.}
\newcommand{\jgeores}{J. Geophys. Res.}
\newcommand{\prd}{Phys. Rev. D}
\newcommand{\aap}{A\&A}

\maketitle

\label{firstpage}

\begin{abstract}
We present a new algorithm to rapidly and optimally compute power
spectra.  This new algorithm is based on a generalization of iterative
multigrid, and has computational cost ${\cal O}(N \log N)$, compared
to the standard brute force approach which costs ${\cal O}(N^3)$.  The
procedure retains this speed on the full sky and for ill-conditioned
matrices.  It is applicable to galaxy power spectra, CMB, polarization
and weak lensing data.  We present a mathematical
convergence analysis, and performance results.
\end{abstract}
\begin{keywords}
statistics, CMB, large-scale-structure, gravitational lensing
\end{keywords}

\section{Introduction}

Cosmological data sets are rapidly becoming very large, making the
optimal analysis of data a challenging problem.  Particularly large
problems arise in the analysis of CMB data, galaxy power spectra and
weak lensing data.  For Gaussian random fields, the statistics are
fully solvable, and the construction of optimal quadratic or maximum
likelihood estimators is uniquely defined.

A full maximum likelihood analysis of COBE which contained 3890 pixels
was done by \citet{1997ApJ...480....6B}.  The optimal analysis
procedure requires ${\cal O}(N^3)$ operations for $N$ data points,
which is generally very expensive, and prohibitive for many data sets.
Data compression is possible through the use of a Karhunen-Loeve
expansion in terms of ``signal-to-noise'' eigenmodes
\citep{1995PhRvL..74.4369B}.  The first compression step still costs
$O(N^3)$, but subsequent analyses are accelerated.  Faster procedures
have been suggested, usually at the expense of optimality.  The
computation of correlation functions, for example, only costs ${\cal
O}(N\log N)$ operations, and has been a popular quick algorithm
\citep{2002ApJ...567...31P,2001ApJ...548L.115S}.  Similarly, a
weighted Fourier Transform of the data is equally fast, and was
proposed by \citet{1994ApJ...426...23F}.  \citet{2003astro.ph..2031P}
summarized the trade-off made in these approximations.

We consider a power spectrum estimator optimal if it minimizes the
variance.  The variance of the power is a four point function, which
can be expressed in terms of two point functions for a Gaussian random
field.  One can show that a naive quadratic estimator applied to the
Wiener filtered data set is optimal \citep{1998ApJ...506...64S}.  If
the noise and signal covariance matrices commute, the correlation
function and weighted Fourier techniques are optimal.  If the
signal-to-noise is less than unity on all scales, the correlation
function is optimal.  But in general none of these cases applies, in
which case the historic fast estimators are suboptimal.

In this paper we present an algorithm which is fast and optimal, and
works on a wide range of problems, geometries, and condition numbers.
It is an extension of the iterative technique presented in
\cite{2002astro.ph.10478P} to cover regimes of poor condition number.

The basic premise is that the forward operation of applying the signal
and noise correlations to the data set an be done in ${\cal O} (N\log
N)$ time.  We then construct iterative techniques to achieve the
inverse.

\section{Maximum Likelihood}

In this section we review the mathematical formulation of the power
spectrum estimation problem.  An observation is a data set of $N$
points sampled at positions $x_i$, written as a vector $\Delta_i$.  We
take this vector to have zero expectation value.  This vector is
assumed to be the sum of signal plus noise, such that the expectation
of the covariance matrix is
\newcommand{\bC}{{\bf C}}
\newcommand{\bE}{{\bf E}}
\beq
\langle \Delta_i \Delta_j \rangle = C^N_{ij}+C^S_{ij}
\eeq
for a noise covariance matrix $\bC^N$ and a signal covariance $\bC^S$.
We parametrize the signal covariance matrix as a sum of $n_b$ band powers,
\beq
\bC^S=\sum_{\alpha=1}^{n_b} p_\alpha \bC_\alpha.
\eeq
The log of the probability of the data given a model is
\beq
{\cal L}=-\frac{1}{2} \log |\bC^N+\bC^S|-\frac{\Delta^T (\bC^N+\bC^S)^{-1}
\Delta}{2} + \frac{n_b}{2}\log(2\pi)
\label{eqn:l}
\eeq
We wish to find the set of parameters $p_\alpha$ which maximize the
likelihood (\ref{eqn:l}).  Differentiating, one obtains a system of
equations
\beq
-\Tr [(\bC^N+\bC^S)^{-1}\bC_\alpha ] + \Delta^t
(\bC^N+\bC^S)^{-1}\bC_\alpha(\bC^N+\bC^S)^{-1} \Delta =0.
\label{eqn:ml}
\eeq
The Newton-Raphson solution for (\ref{eqn:ml}) starts with an initial
guess $p_\alpha^0$ and iterates
\newcommand{\bF}{{\bf F}}
\newcommand{\bG}{{\bf G}}
\beq
p_\beta^{n+1}=p_\beta^n-f_\alpha G^{-1}_{\alpha\beta}
\label{eqn:newton}
\eeq
where $f_\alpha$ is the left hand side of (\ref{eqn:ml}) and $\bG$ is
the Jacobian
\beq
G_{\alpha\beta}=2F_{\alpha\beta} - \Delta^t
(\bC^N+\bC^S)^{-1}\bC_\alpha(\bC^N+\bC^S)^{-1}\bC_\beta(\bC^N+\bC^S)^{-1}
\Delta
- \Delta^t
(\bC^N+\bC^S)^{-1}\bC_\beta(\bC^N+\bC^S)^{-1}\bC_\alpha(\bC^N+\bC^S)^{-1}
\Delta.
\label{eqn:jacobian}
\eeq
$\bF$ is the Fisher matrix defined as
\beq
F_{\alpha\beta}\equiv\frac{1}{2}\Tr[ (\bC^N+\bC^S)^{-1}\bC_\beta
(\bC^N+\bC^S)^{-1}\bC_\alpha ].
\eeq

Newton-Raphson only convergences in the domain where the sign of the
curvature does not change.  But even for one degree of freedom, the
Gaussian likelihood encounters an inflection point.  Consider
\beq
2\log {\cal L}=-\log(2\pi)-2\log(\sigma)-\frac{\Delta^2}{\sigma^2}.
\label{eqn:l1d}
\eeq
We solve the equation
\beq
f(\sigma)\equiv\frac{1}{\sigma}-\frac{\Delta^2}{\sigma^3}=0.
\label{eqn:root1}
\eeq
The solution is $\sigma^2=\Delta^2$, as one would expect.
In one dimension the Jacobian is the derivative,
$f'(\sigma)=-1/\sigma^2+3\Delta^2/\sigma^4$, which changes sign when
$\sigma^2=3\Delta^2$.  So the Newton-Raphson iteration
(\ref{eqn:newton}) moves towards the correct root for guesses within a
factor of 3 the true solution, but moves towards infinity for guesses
that are too large.

\newcommand{\bW}{{\bf W}}
An alternate representation is in terms of weighted quadratic
estimators, where one first weights data by $\bW$
and defines raw estimators $q_\alpha$ as
\beq
q_\alpha \equiv \Delta^t
\bW^t\bC_\alpha\bW \Delta - \Tr [ \bW \bC_\alpha \bW C_N ] .
\label{eqn:quadratic}
\eeq
The estimators of power $\tilde{p}_\alpha=F^{-1}_{\alpha\beta}q_\beta$.
A Wiener filter choice is $\bW=(\bC^N+\bC^S)^{-1}$.  If interpreted in
terms of the maximum likelihood iteration (\ref{eqn:newton}), it
amounts to an initial guess $p_\beta^0=0$, with residual
$f_\alpha=q_\alpha$.  We used the expectation value $\langle \bG \rangle=\bF$.
This converges after a single iteration.  Of
course, we had assumed that we already knew the optimal weights for
the Wiener filter.  The quadratic estimator (\ref{eqn:quadratic}) does
not have convergence limitations.  One can start with an approximate
guess for the covariance $\bC^S$, and iterate.

In both the maximum likelihood and the quadratic estimation processes,
a number of expensive operations are required.  The most common one is
the solution to the Wiener filter, $(\bC^N+\bC^S)^{-1}\Delta$.  We
will discuss in section \ref{sec:iter} how to achieve using a fast
algorithm, after which we only need to address the trace evaluation,
which we do in section \ref{sec:trace}.

\section{Iterative Solution}
\label{sec:iter}

For stationary processes, the correlation matrices depend only on the
separation distance between points.  This makes them diagonal in
Fourier space.  Noise matrices are often diagonal in real space, or in
a time stream space.  For now we assume that a fast ${\cal O}(N log N)$
method exists to evaluate the forward operation.  In section
\ref{sec:implement} we present several fast procedures which imlement
this.

For concreteness, we assume that the signal correlation is a
diagonal matrix in Fourier space, while the noise matrix is diagonal
in real space.  We concentrate on the problem
\newcommand{\bN}{{\bf N}}
\newcommand{\bI}{{\bf I}}
\newcommand{\bS}{{\bf S}}
\newcommand{\bH}{{\bf H}}
\newcommand{\bL}{{\bf L}}
\newcommand{\brS}{\bar{S}}
\beq
({\bf S+N}) x = y .
\label{eqn:sn}
\eeq
The first requirement is that (\ref{eqn:sn}) can be evaluated
quickly.  This is demonstrated for a range of problems in section
\ref{sec:convolve}.
We will express all vectors in real space.  We can evaluate
$y$ in ${\cal O}(N \log(N))$ operations as $y={\bf N} x + {\bf
S} x$.   The underrelaxed Jacobi iterative solution is
\beq
x^{n+1}=x^n+ (\bar{S} \bI+\bN)^{-1} \left[y- (\bS+\bN) x^n\right].
\label{eqn:jacobi}
\eeq
$\bar{S}$ is our relaxation parameter, to be determined below.
Define the {\it error} of the $n-$th iteration as the difference
between the true solution $x$ and the current guess, $e^n=x^n-x$.
Substituting into (\ref{eqn:jacobi}), we have
\beq
e^n=\left[\bI-(\bar{S}\bI+\bN)^{-1} ({\bf S+N})\right]^n e^0 \equiv
\bE^n e^0.
\label{eqn:err}
\eeq
$\bE$ is called the {\it error matrix}.
We define the {\it spectral radius} of a matrix ${\bf M}$ as
$\rho({\bf M})$ to be the maximum absolute eigenvalue of ${\bf M}$.  
Since $\rho({\bf A+B}) \leq \rho({\bf A})+\rho(\bf{B})$,
the sufficient condition for (\ref{eqn:err}) to converge is 
\beq
\bar{S}\ge\rho(\bS).
\label{eqn:conv}
\eeq
We can prove (\ref{eqn:conv}) by expanding the matrix product in
(\ref{eqn:err}) into two positive definite terms each with spectral
radius less than unity, $\rho[(\bar{S}\bI+\bN)^{-1} \bN]<1$ and
$\rho[(\bar{S}\bI+\bN)^{-1} \bS]<1$.  The first inequality is trivial
since the two terms are simultaneously diagonal.  The second
inequality follows by expanding its inverse, and noting that the
smallest eigenvalue of the inverse is larger than one.

When $\bN$ is diagonal, we obtain optimal convergence with
$\bar{S}=(S_0+S_\infty)/2$, i.e. the average of the largest and
smallest eigenvalue of the noise matrix.  For practical purposes, we
have never encountered convergence problems with this choice, and
always found better convergence than using the hard bound
(\ref{eqn:conv}). 

In real space, (\ref{eqn:err}) can be visualized as follows.  We take
an error vector $e^n$, multiply one copy by the pixel noise, and
convolve another copy with the two point correlation function.  We add
the two resulting vectors, and divide by the pixel noise plus
a constant corresponding to the amplification factor for the wave with
the largest signal power.  We then subtract this from the original
vector $e^n$.  The convergence is rapid if it pointwise converges to
zero.  The pixels which are slowest to converge are those where the
noise is much smaller than $\bar{S}$ when we inject a wave
corresponding to the smallest eigenvalue of the signal correlation.
The worst pixel converges as 
\beq
|e^n|_\infty \propto [1-(N_0+S_0)/(N_0+\bar{S})]^n
\label{eqn:e0}
\eeq
where $N_0$ is the smallest eigenvalue of $\bN$, $S_0$ is the
smallest eigenvalue of $\bS$, and $\bar{S}$ is the largest eigenvalue. 
It takes $\sim (N_0+\bar{S})/(N_0+S_0)$ steps to decrease the error by
an $e$-fold.  The rate is related to the condition number of $\bS$,
and the lower limit for $\bN$.
It is apparent that choosing $\brS$ larger than necessary slows the
convergence down.

The convergence rate is limited by the condition number of $\bS+\bN$.
Consider the problem of power spectrum estimation on an irregular region with
constant noise.  We can express this problem on a larger enclosing
domain, and write $N^{-1}_{ii}=0$ on the padding region.  $\bN^{-1}$ is not
invertible in this case.
But we can write
\beq
(\bS \bN^{-1}+\bI) \bN x=y.
\label{eqn:sni}
\eeq
Setting $u=\bN x$, we first solve (\ref{eqn:sni}) for $u$, and then
evaluate $x=\bN^{-1} u$, where zero entries on the diagonal of
$\bN^{-1}$ becomes a non-problem.  The problem instead arises when the
noise is small.  Large and small is defined relative to the signal.
Since the signal covers a spectrum of values, we shall use the mean
value $S_M$ as reference point.  We factor $\bN=\bH \bL$, where
$\bH=\bN$ when $\bN>S_M$ and equal to one otherwise.  We can now
generalize (\ref{eqn:sni}) to
\beq
(\bS \bH^{-1}+\bL) \bH x=y,
\label{eqn:snhl}
\eeq
and define $u=\bH x$.  The iterative solution becomes 
\beq
u^{n+1}=u^n+ (\bar{S}\bH^{-1}+\bL)^{-1} \left[y- (\bS\bH^{-1}+\bL) u^n\right].
\label{eqn:j2}
\eeq
The convergence criterion (\ref{eqn:conv}) is still a sufficient
condition for the Jacobi iteration (\ref{eqn:j2}), but may be quite
non-optimal. We can instead use an iterative estimate of
$\bar{S}=\rho(\bS\bH^{-1})$ using a power iteration.
To prove the convergence of (\ref{eqn:j2}), we use the same argument
as for (\ref{eqn:conv}), with the second inequality holding since
$\bH\bS^{-1}\bH^{-1}$ is a similarity transformation on $\bS$ which
preserves eigenvalues.
\newcommand{\bU}{{\bf U}}
\newcommand{\bT}{{\bf T}}
Should $\bS$ have a large dynamic range, we can similarly factor it
$\bS=\bT \bU$ 
\beq
(\bT \bH^{-1}+\bU^{-1}\bL) \bH x=\bU^{-1}y,
\label{eqn:snbu}
\eeq
with an iteration
\beq
u^{n+1}=u^n+ (\bar{T}\bH^{-1}+\bar{U}^{-1}\bL)^{-1} 
\left[y- (\bT\bH^{-1}+\bU^{-1}\bL) u^n\right].
\label{eqn:j3}
\eeq
A similar argument shows this to be a convergent iteration if we
choose $\bar{T}=\rho(T)$ and $\bar{U}=1/\rho(\bU^{-1})$.

The iterative approach also allows us to compute the $\chi^2=\Delta^T
(\bC^N+\bC^S)^{-1} \Delta$ of the solution iteratively, as well as any
expression involving the contraction of vectors, such as
(\ref{eqn:quadratic}).  The likelihood itself requires computing a
determinant, which does not fall into this framework.  The trace
evaluations needed in Equations (\ref{eqn:ml},\ref{eqn:jacobian}) will be
addressed below.

\section{Convolution and boundaries}
\label{sec:convolve}

\subsection{Multi-level FFT}

Consider a point set contained within an irregular domain $\Omega$
with isolated boundary conditions.  We imbed this domain $\Omega$ in
the smallest enclosing rectangle $L$.  The multiplication by the
signal matrix is just a convolution by the correlation function with
isolated boundary conditions, which can be performed using an FFT on a
domain twice as wide as $L$.  The operation $y=\bS x$ sets the value
at every point $y_i$ to the sum over all points $x$ weighted by the
correlation function at the separation.  We can map any discrete set
of points onto a regular grid using an interpolation of some order
\citep{1988csup.book.....H}. 

For this iterative technique, we only rely on each of the forward
operations to be evaluated rapidly.  The operation of the signal on
the data can always be performed in ${\cal O}(N \log N)$ time on a
spatial grid by considering it a convolution, which can be performed
on a tree, even if the geometry is complex, the coordinate system is
unevenly sampled, or the coordinates are otherwise not appropriate for
a Fourier transform, for example on the celestial sphere.  Again,
several methods are possible, with tradeoffs in complexity of the
algorithm against computational time.  As discussed in
\citet{2002astro.ph.10478P}, several levels of grids can be employed
to accelerate the process.  Open boundaries can be implemented, and
one can add short range pair summations, much in analogy to N-body
codes \citep{1988csup.book.....H}.  If correlations on a whole sphere
are desired, a spherical harmonic transform can be applied for
convolutions on the coarse grid followed by FFT's on the fine grids.
While spherical harmonic transforms cost ${\cal O}(N^{3/2})$, the base
is only the number of coarse grid cells, and one expects to be
dominated by the ${\cal O}(N\log N)$ of the fine grid in general.

\subsection{Tree}
\newcommand{\vx}{\vec{x}}

Just as in N-body simulations, the optimal algorithm depends on the
clustering of points.
An alternative way of evaluating the signal correlation applied to a
data vector $\bS \Delta$ is through tree summation.  This approach is
very general, and does not slow down under strong clustering.
A stationary signal correlation is a convolution, and we can express it as
\beq
u(\vx_i)\equiv \bS \Delta = \sum_j \Delta(\vx_j) G(|\vx_i-\vx_j|).
\label{eqn:convolve}
\eeq
While Equation (\ref{eqn:convolve}) is an $N^2$ process, we note that it can be
approximated to high accuracy through a multipole expansion.  Points
can be bunched together into nodes.  Each node has the mean position
of its constituent points (center of mass), a quadrupole moment, and
the sum the masses.  So instead of summing over all points, it is
sufficient to sum over a smaller number of nodes.  Let us describe the
mass distribution of node $k$ by $\rho_k(\vx)$.  The convolution over
the node is
\begin{eqnarray}
u(\vx_i)&\equiv &\bS \Delta = \sum_k \int dx \rho_k(\vx)
G(|\vx_i-\vx|)
\nonumber\\
&=& \sum_k \left[ M_k G(|\vx_i-\vx_k|) + \int \rho_k(\vx) r^\alpha
r^\beta \partial_\alpha \partial_\beta G(|\vx_i-\vx_k|) + {\cal
O}(G''')\right].
\label{eqn:taylor}
\end{eqnarray}
The mass $M_k=\int \rho_k$, and
dipole term was cancelled by the choice $\langle \vx_k\rangle
=\frac{1}{M_k}\int \vx \rho_k(\vx)$, and  The second
term in (\ref{eqn:taylor}) can be rewritten in terms of the quadrupole
moments $Q_{\alpha\beta}=\int r^\alpha r^\beta \rho_k$ with
$r^\alpha=\vx^\alpha_k-\vx^\alpha$ as
\beq
u(\vx_i)
= \sum_k \left\{ M_k G(|\vx_i-\vx_k|) + Q_{\alpha\beta}^k
\left[(\delta_{\alpha\beta} - \hat{r}^\alpha\hat{r}^\beta )
\frac{G'(|\vx_i-\vx_k|)}{|\vx_i-\vx_k|}
+\hat{r}^\alpha\hat{r}^\beta 
G''(|\vx_i-\vx_k|) \right] + {\cal  
O}(G''')\right\}.
\label{eqn:quad}
\eeq
$\hat{r}\equiv \vec{r}/|r|$ is the unit separation between the target
point and the node center of mass.
By comparing the quadrupole to the monopole, we obtain an estimate of
the error in the truncation.  Should this error be larger than our
tolerance, we break the sum into subcomponents of the node.

The tree needs only be constructed once, so we should invest
sufficient computing resources into this step to minimize the
truncation errors arising from the tree.  One starts by defining a top
node, which includes all points.  Since our goal is to maximize
locality of particles, we shall subdivide each node by cutting it in
half by a line perpendicular to the major axis of the quadrupole.  We
want to place the cut such that the quadrupole of the subnodes is
minimized.  This cut line can be determined through bisection.  One
first cuts through the center of mass, and determines the quadrupole
of each subnode.  We then displace the cutline towards the node with
the larger quadrupole.  Let $r_m=\sqrt{\lambda_{\rm max}}$ be the
square root of the largest eigenvalue of the parent node.  We then
displace the cutting line by an amount proportional to $r_m$ times 1
minus the ratio of quadrupoles of the subnodes.  We repeat this
process on each subnode until each node contains exactly one point.
The resulting binary tree is not balanced, since the number of
particles in each child node is not equal.  The depth of the tree is
not determined in advance.

The truncation error in the quadrupole tree for a power law correlation
function $G(R)\propto R^{-n}$ is
\beq
e\propto \frac{r_m^3 G'''(R)}{G(R)}= n(n+1)(n+2) \left(\frac{r_m}{R}\right)^3.
\eeq
Reducing $r_m$ by a factor of two reduces the error by a factor of 8.  But
the quadrupole is proportional to the node area, so the computational
cost has increased a factor of 4.  Let $n_2$ be the number of desired
binary digits of accuracy.  Then the quadrupole truncated tree has a
computational cost $\propto (2^{n_2})^{2/3}$.  If very high accuracy is
desired, one should go to high multipole order.  An order $n_O$ multipole
evaluation in two spatial dimensions costs ${\cal O}(n_O)$ operations,
so the total node cost $\propto n_O^2$.  At high multipole moments,
it makes sense to reduce the node size $r_m$ since we gain rapidly.
The optimal $n_O$ is thus slightly smaller than $n_2$.
Asymptotically, the cost of the tree slightly less than $n_2^2$.

A simple tree decomposition to implement does not use the quadrupole
information.  For each node, we find the particle which is furthest
from the center of mass.  We store this distance with each node.  We
subdivide the node into two subnodes by a cut perpendicular to the
line connecting the furthest particle to the center of mass.  The cut
is chosen such that the two subnodes have an equal number of
particles.  When walking the tree, we check that the maximal radius of
the node is sufficiently smaller than the distance to the node.  This
tree is faster to build, since no quadrupole evaluations are required,
and for each node the bisection only costs linear time in the number
of particles in the node.  It also minimizes the worst case error at
each node.

\section{Multiscale Acceleration}

\newcommand{\bPi}{{\bf P}_i}
\newcommand{\bP}{{\bf P}}
The convergence rate of equation (\ref{eqn:j2}) is primarily limited
by the condition number of $\bS$, since the relaxation parameter
$\bar{S}$ is yields rapid convergence on modes with correlation power
close to $\bar{S}$.  We can, for example, project the error ${\bf e}$
onto the eigenvectors of $\bS$.  Since we chose $\bar{S}$ to be the
largest eigenvalue of $\bS$, the corresponding eigenvector converges
to zero error after one iteration.  We thus consider breaking the
iterative solution into blocks of the eigenspace of $\bS$, for which
the corresponding eigenvalues have a small dynamic range.  If we can
use a different relaxation parameter for each block, the iteration
proceeds rapidly.  Let us write a complete set of $n_b$ band power
projection 
operators $\bPi$ which sum to the identity matrix.  
Multiplication by the projection can also be expressed as a
convolution, for which the tree algorithm described above can
also be used.
We generalize the
iteration from equation (\ref{eqn:j3})
\beq
u^{n+1}=u^n+ \sum_i(\bar{T}_i\bH^{-1}+\bar{U}_i^{-1}\bL)^{-1} 
\bPi\left[y- (\bT\bH^{-1}+\bU^{-1}\bL) u^n\right].
\label{eqn:mg}
\eeq
The error on the iteration is again
\beq
e^n=\left[\bI-\sum_i(\bar{T}_i\bH^{-1}+\bar{U}_i^{-1}\bL)^{-1} 
\bPi(\bT\bH^{-1}+\bU^{-1}\bL) \right]^n e^0.
\label{eqn:errmg}
\eeq
We set the relaxation parameters $\bar{T}_i=\rho(P_i \bT)+T_\infty$
and $\bar{U}_i=1/\rho(P_i \bU^{-1})+U_\infty$.  In the continuous
Fourier decomposition limit, we have
\newcommand{\bk}{{\bf k}}
\newcommand{\bx}{{\bf x}}
\beq
e^{n+1}(x)=e^n(x)-\int d^2k d^2x' \exp[i\bk\cdot(\bx-\bx')] 
\frac{P(\bk)+N(\bx')}{P(\bk)+N(\bx)} e^n(\bx').
\label{eqn:cerr}
\eeq
The error matrix (\ref{eqn:cerr}) is zero on the diagonal.  If there
is no signal, $P(k)=0$, the error matrix is zero everywhere, and the
convergence is exact after one iteration.  When the signal is
dominant, we can neglect $N(x)$, and the error matrix is again zero.
Similarly, the error matrix is zero for two elements whenever
$N(x')=N(x)$.

We can solve the error matrix exactly for two degrees of freedom.  In
this case, we write the noise matrix as 
\beq
\bN=\left(\begin{array}{cc} n_1 &  0 \\
			     0  & n_2   \end{array}\right) .
\eeq
The orthogonal Fourier transform matrix $\bF$ is
\beq
\bF=\frac{1}{\sqrt{2}}\left(\begin{array}{cc} 1 &  1 \\
			                      1  & -1
\end{array}\right) .
\eeq
We have two projection matrices
\beq
\bP_1=\left(\begin{array}{cc} 1 &  0 \\
	                      0 & 0
\end{array}\right) , \ \ \ \ \ 
\bP_2=\left(\begin{array}{cc} 0 &  0 \\
	                      0 & 1
\end{array}\right).
\eeq
The signal correlation matrix is
\beq
\bS=\bF^{-1}\left(\begin{array}{cc} s_1 &  0 \\
			     0  & s_2   \end{array}\right) \bF.
\eeq
The error expression is then
\beq
e^n=\left\{\bI-[((s_1+s_\infty)\bI+\bN)^{-1}\bF^{-1}\bP_1
+((s_2+s_\infty)\bI+\bN)^{-1}\bF^{-1}\bP_2]  
(\bS \bF+\bF \bN)\right\}^ne^0
\label{eqn:2it}
\eeq
For $s_\infty=0$, the eigenvalues of this expression can be directly
evaluated
\beq
\lambda=\pm i \frac{(n_1-n_2)(s_1-s_2)}{
\sqrt{(n_1+s_1)(n_2+s_2)(n_1+s_2)(n_2+s_1)}}.
\label{eqn:lsn}
\eeq
If we take $n_i$ and $s_i$ to be sorted positively in ascending order,
we see the qualitative features of the convergence.  Convergence is
very rapid if either $\bN$ or $\bS$ is well conditioned.  It is poor
if $n_1+s_1$ is very small compared to all the other differences in
the problem, and can even fail to converge if $|\lambda|>1$.
To treat the most extreme case, we set $s_1=n_1=0$.  Then the
iteration (\ref{eqn:2it}) will converge optimally for $s_\infty \sim {\rm
min}(s_2,n_2)$, and the maximal eigenvalue is $\lambda<1/2$ for any
choice of $n_2,s_2$.  At least in this scenario, we have shown that
even this potentially extremely ill-conditioned system converges
rapidly, better than a factor of two on each iteration.

In the continuum limit (\ref{eqn:cerr}), we can estimate convergence
in some limiting cases.  If $N(x')\gg P(k)$ is large the $k$ integral
turns the power spectrum into a correlation function,
$\xi(\bx'-\bx)/N(\bx')$, while if it is very small, the matrix element
is given by the inverse correlation function $N(\bx)
\xi^{-1}(\bx-\bx')$.  In either case, the matrix is very small.

When the signal matrix contains a large number of zero entries,
setting $T_\infty$ to the mean value of the noise appears to be a
robust choice.  In general, we break the sum (\ref{eqn:errmg}) into
$\log_2(||\bS||)$ terms, each of which costs ${\cal O}(N\log N)$
operations.

\begin{figure}
\vskip 3.3 truein
\includegraphics{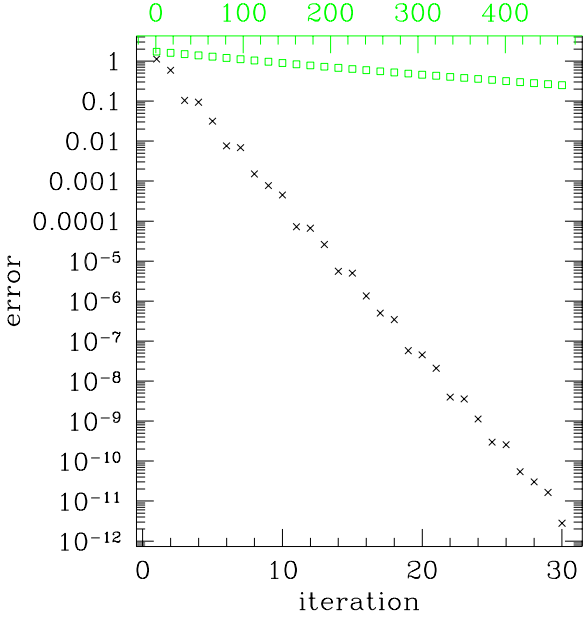}
\caption{Convergence of Jacobi iteration and multiscale acceleration.
Bottom horizontal label is the number of the iteration for the
multiscale accelerated method, while the top axis is the iteration
number for direct Jacobi iteration.  The problem had $65536$ variables.
}
\label{fig:err}
\end{figure}

To test these concepts, we implemented a one dimensional example.
65536 grid points were used with periodic boundary conditions, and a power
law signal correlation function $P(k)=256/k$, where $k$ is the integer
wave number on the grid.  The noise variance in
each grid point is a random number chosen uniformly from 0.35 to 1.35.
We first run the straight iteration (\ref{eqn:jacobi}) for 500
iterations.  The vector $\vec{x}$ was chosen randomly, from which
$\vec{y}$ was computed.  We define the scaled infinity error
$e_\infty=|x^n-x|_\infty\sqrt{65536/\sum x_i^2}$.  This error is plotted as
boxes in figure \ref{fig:err}.
In this situation, we have $\bar{S}\sim 128$, $S_0=1/256$ and
$N_0=0.35$.  From equation (\ref{eqn:e0}) we expect to take up to about
366 iterations to reduce the error by an $e$-fold.  This is consistent
with the performance seen in figure \ref{fig:err}.

Then we applied the multiscale accelerated iteration (\ref{eqn:mg}).
We chose 16 bands, so the computational cost is about 16 times higher,
but the convergence is much better as we see in the crosses in figure
\ref{fig:err}.  The two sets of points have comparable computational
cost, and the multiscale converges much better.  We note that the
convergence rate for the Jacobi iteration scales as the condition
number, which is proportionate to the number of grid points, while the
multiscale convergence is independent of that.  The Jacobi iteration
is still very fast compared to a brute force $N^3$ solution of this
problem.  Each of the iterative methods here take a matter of minutes
on a workstation, while a brute force method takes of order a year.

\newcommand{\bD}{{\bf D}}
\newcommand{\cR}{{\cal R}}
\newcommand{\cP}{{\cal P}}
\newcommand{\tbL}{\tilde{\bL}}

It is instructive to compare this multiscale approach to traditional
multigrid aceleration methods.  We solve an arbitrary linear equation
\beq
\bL u=b
\eeq
for some linear operator $\bL$.  The simplest procedure applies a
Jacobi iteration
\beq
u^{n+1/2}=\left(1-\bD^{-1}\bL\right) u^n+\bD^{-1}b,
\eeq
where $\bD$ is the diagonal part of $\bL$.
We then have a {\it restriction operator} $\cR$ which maps the vector
$u$ onto a coarser vector $u_1=\cR u$.  In this coarser space, we need
to have a coarsened version of the operator, denoted $\tbL$, for which
we now solve the coarsened system
\beq
\tbL u_1^{n+1/2}=\Delta b_1^{n+1} \equiv \cR(b-\bL u^{n+1/2}).
\label{eqn:coarse}
\eeq
One recursively applies multigrid to this coarsened system.  For
purposes of analysis, we assume that the coarse grid equation
(\ref{eqn:coarse}) was solved exactly.  We inject this coarse grid
correction using the coarse-to-fine grid prolongation operator $\cP$
\begin{eqnarray}
u^{n+1}&=&\cP \tbL^{-1} \Delta b^{n+1/2}+u^{n+1/2}
\nonumber\\
&=&u^n
+  \left[\cP \tbL^{-1}\cR(1-\bD^{-1}\bL)+\bD^{-1}\right] (b-\bL u^n).
\end{eqnarray}
The error can be written as
\beq
e^n=\left[(1-\cP \tbL^{-1}\cR\bL)(1-\bD^{-1}\bL)\right]^n e^0.
\label{eqn:erromg}
\eeq
The standard Jacobi iteration only has the right term in parenthesis.
For the Laplace operator $\bL=\nabla^2=-k^2$, we can see in Fourier
space that this term is dominated by small $k$, i.e. large scales, for
which the term is unity, and convergence is slow.  But on large
scales, the restriction and prolongation operators basically commute
with everything, and the first term is very small.  So we see that
convergence is rapid on all scales for the Laplace operator.

In correlation problems, the linear operator is typically an inverse
power of the wave number, $L\sim k^{-n}$, for which the large scales
actually converse rapidly, but the small scales don't.  On these small
scales, the prolongation and restriction operators introduce large
errors, and the convergence is not aided by this multigrid
decomposition.  One can remedy this problem by multiplying both sides
of the equation by a sufficiently large power of $k$.
Or one can recast the equation (\ref{eqn:sn}) as
$({\bS^{-1}+\bN^{-1}}) \bN x = \bS^{-1} y$, and solve the system using
only inverse matrices.  But as we had
seen in equations (\ref{eqn:e0},\ref{eqn:lsn}), the convergence is
generally limited by a bad condition number on the noise matrix
$\bN$.  Let us consider the above 2x2 example, choosing the case when
$n_1=s_1=0$.  For the coarse grid, the only conceivable operator is
the scalar $\tbL=(s_2+n_2)/2$.  The restriction to a one cell coarse
grid is the average of the two cells, and vice versa.
The iterator error (\ref{eqn:erromg})
is then
\beq
e^n=\bE^n e^0 \equiv \left[ \left( \begin{array}{cc}
	1 & -\frac{n_2}{s_2+n_2} \\
	0 & \frac{s_2}{s_2+n_2} \end{array} \right)
\left( \begin{array}{cc}
	0 & -1 \\
	- \frac{s_2}{s_2+2n_2}& 0 \end{array} \right)\right]^n e^0.
\eeq
The term in the right parentheses is the error for a standard Jacobi
iteration without multigrid, and the left term is the multigrid
acceleration.  In a pure Jacobi iteration, the eigenvalues are
$\lambda=\pm \sqrt{\frac{s_2}{s_2+2n_2}}$, which tends to 1 as $n_2$
tends to zero and is thus slowly convergent in some cases.  The
combined error matrix has eigenvalues $\lambda=\frac{s_2(n_2\pm
i\sqrt{3n_2+2s_2}\sqrt{5n_2+2s_2})}{2(n_2+s_2)(2n_2+s_2)}$, which are
still always less than one, but approach 1 as $n_2$ approaches zero.
The full multigrid scheme thus does not solve the illconditioning
problem of the noise matrix.  As we showed above, the generalized
multiscale decomposition proposed here (\ref{eqn:2it}) is always well
conditioned and converges by a factor of 2 every iteration for any
choice of signal or noise.  The computational cost of the multiscale
method is logarithmically more expensive.

In practice, maps often has ragged edges, defect, and other excisions,
which generically give rise to large entries in the noise matrix, and
make it ill-conditioned.  For polarization maps, large regions of the
signal matrix may also be very small (e.g. the $B$-mode discussed
below).   Multigrid methods must bin pixels with widely varying noise,
which generically gives rise to convergence problems.  The prologation
of the inverse of the coarse grid operator $\tbL^{-1}$ does not
necessarily cancel the fine grid operator $\bL$ on coarse scales.  We
conclude that the multiscale procedure proposed here is more general
than multigrid.

\section{Trace Evaluation}
\label{sec:trace}

Evaluation of Equation (\ref{eqn:ml}) requires not only solving linear
equations, but also the evaluation of a trace.  
We propose a rapid
evaluation similar to that used in \citet{1999ApJ...510..551O}.
Consider a vector $v_i$ with random elements 
which are either +1 or -1.  Then
\beq
\langle v_i v_j \rangle = \delta_{ij}.
\eeq
Thus,
\beq
\langle v^T(\bC^N+\bC^S)^{-1}\bC_\alpha v \rangle = 
\Tr[(\bC^N+\bC^S)^{-1}\bC_\alpha] .
\label{eqn:vt}
\eeq
The vector equation on the left only requires solving linear systems,
which we can evaluate rapidly.

If we now construct a series of orthogonal vectors $v$, then an
average over $N$ vectors makes (\ref{eqn:vt}) exact.  We construct an
orthogonal sequence by setting the first vector to be +1 for the first
half elements, and -1 for the second half, the second vector +1 for
the first and third quartile, and -1 on the remainder, etc.  We then
apply a random shuffle permutation on each vector, which results in a
random orthogonal sequence.  Each of these operations is ${\cal O}(N)$.

For a diagonal matrix, a single iteration is always exact.  For a
positive definite matrix, the convergence rate is approximately
\beq
\frac{|\delta \Tr|}{\Tr} \sim \frac{1}{\sqrt{n}}\frac{\left(\sum
\lambda_i^2 \right)^{1/2} }{\Tr}
\label{eqn:errtrace}
\eeq
where $N$ is the linear size of the matrix, and $n$ is the number of
iterations.  If the eigenvalues are of comparable magnitude, the RHS
of (\ref{eqn:err}) is $\propto 1/\sqrt{Nn}$.  In the worst case, all
eigenvalues except for one is zero, in which case the convergence is
$\propto 1/\sqrt{n}$.

When we replace the trace by a stochastic estimate as done in equation
(\ref{eqn:vt}), the corresponding Jacobian (\ref{eqn:jacobian}) in the
maximum likelihood solution is no longer symmetric.  The strategy to
test for the sign of the eigenvalues may fail in this case, since
eigenvalues are no longer guaranteed to be real.  Empirically, we find
that checking the eigenvalues of the symmetrized Jacobian works as a
good indicator of convergence problems.  When the sign of the
eigenvalues is incorrect, we use the symmetrized eigensystem to
correct the iteration direction.  This symmetrized jacobian only gives
first order convergence, instead of the second order expected from the
exact Newton-Raphson.  So we simply switch to the standard iteration
in the vicinity of the solution, and find rapid convergence to machine
precision. 

\section{Implementation}
\label{sec:implement}
We will discuss the implementation for weak lensing and galaxy angular
power spectrum analysis.  Statistical weak lensing allows a
measurement of the gravitational field of the dark matter by measuring
the induced alignments of background galaxy shapes.  Most recent
analyses have relied on the correlation function, which is an inverse
noise weighted quadratic estimator.
Computing the two point correlation function is in general not an
optimal estimator of the power.  Being a quadratic estimator, it
weights each data bin by only its local noise.  Each correlation
function bin $q_i$ is an inverse noise $N^{-1}$ weighted estimator of the data
$\Delta$ for a bilinear form $Q_i$:
\begin{equation}
q^i = \Delta^T N^{-1} Q_i N^{-1} \Delta
\end{equation}
while an optimal estimator (\ref{eqn:quadratic}) should have used
$(S+N)^{-1}$ as its 
weights instead.  In the limit that signal to noise is small, the
correlation function is optimal, similarly it is optimal if the signal and
noise covariance matrices commute.  For weak lensing surveys, on small
scales signal to 
noise is small, while on larger scales the coverage is reasonably
uniform, so a correlation function is not a very poor estimator.
As described in \citet{2003astro.ph..2031P}, the correlation function
computation is $O(N \log N)$, so is very fast.

Alternatively, one can grid the data and perform optimal analysis as
described in \citet{2001ApJ...554...67H}.  Due to the large cost
$O(N^3)$, large grid cells must be chosen, leading to non-optimality
in the binning.  The algorithm in this paper combines the advantages
of both approaches.

We can consider each galaxy to have two observables, $g_1,\ g_2$, as
well as a position.  Our goal is to measure the correlations described
as two power spectra.  The noise on each data points tends to be very
large, usually much larger than the signal.  In this regime, the
straight Jacobi iteration (\ref{eqn:jacobi}) has good convergence
according to equation (\ref{eqn:e0}), and we may be able to do
without multiscale acceleration.

In general, the observed galaxy ellipticity is related to the reduced
shear, $\langle g_i\rangle=\gamma_i/(1-\kappa)$, which depends on both
the shear $\gamma$ and the convergence $\kappa$.  In weak lensing,
$\kappa \sim 0$, and we equate the expectation value of the galaxy
alignment with the shear.  The covariance is
\beq
\langle g_i(x_\alpha) g_j(x_\beta)\rangle = c_{ij}(x_\alpha-x_\beta)
+\delta_{\alpha\beta} \delta_{ij}g^2/2.
\eeq
The last term should be the intrinsic ellipticity of the source plane
galaxy.  But that is not directly observable.  The observed ellipticity
is the sum of the actual ellipticity and the lensing induced ellipticity.
In weak lensing, we will ignore the induced ellipticity.  This corresponds
to forcing $c_{ij}(0)=0$, i.e. forcing the power spectrum to have zero
mean.

A generic spin two correlation function $c_{ij}$ can be described in
terms of two power spectra.  The first one, called 'electric', or
'div-like' describes the divergence of the polarization field, $C^E_l$,
and is induced by weak lensing.  The second one, called 'magnetic' or
'curl-like' describes the curl of the polarization field, $C^B_l$.
From these, we construct two correlation functions,
\begin{eqnarray}
c_+(\theta)&=&\frac{1}{2\pi} \int l dl C^E(l) J_0(l\theta)
+C^B(l)J_4(l\theta)
\nonumber \\
c_-(\theta)&=&\frac{1}{2\pi} \int l dl C^E(l) J_0(l\theta)
-C^B(l)J_4(l\theta)
\end{eqnarray}

We can express
\beq
c_{ij}=\left(\begin{array}{cc}
	c_++c_-\cos(4\phi) & c_-\sin(4\phi) \\
	c_-\sin(4\phi) & c_+-c_-\cos(4\phi) \end{array}\right)
\eeq
in terms of the angle $\tan^{-1}\phi=\Delta y/\Delta x$ between
the two galaxy positions.

\begin{figure}
\vskip 6.8 truein
\includegraphics{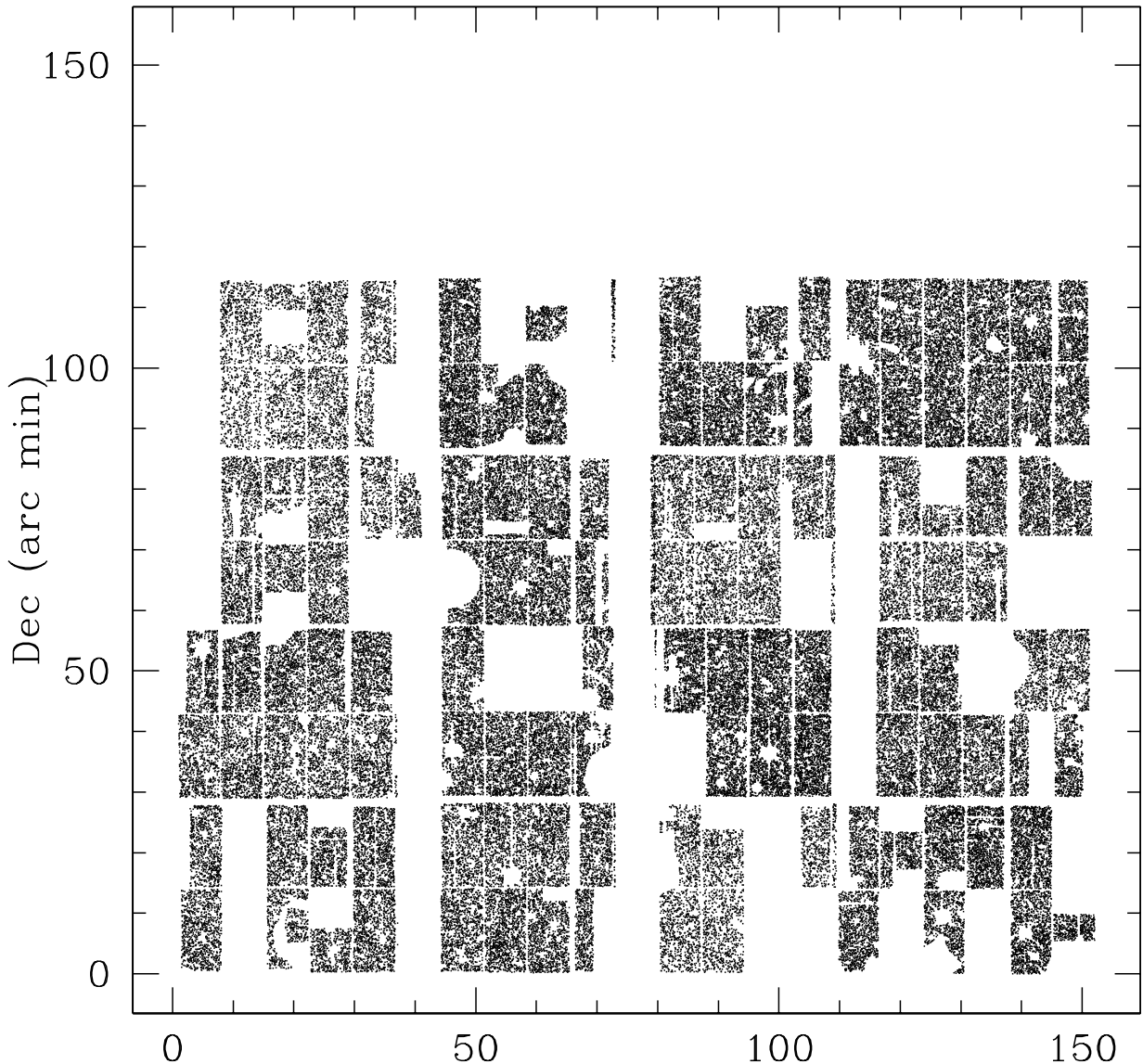}
\caption{Position of $I_{AB}>22$ Virmos galaxies in the 14h field.  The power
  spectrum of dark matter in the form of weak lensing shear
  distortions is sampled at galaxy positions.  This survey geometry is
non-trivial at all angular scales.}
\label{fig:geom}
\end{figure}

To test the procedure, we used a real weak lensing survey geometry as
described by \citet{2003astro.ph..2031P}.  The galaxy positions are
shown in figure \ref{fig:geom}, which corresponds to the survey
geometry.  It is apparent that the geometry is highly irregular, where
the survey geometry contributes significantly more power on all scales
than the intrinsic cosmic signal.  To generate the simulated power
spectrum, we randomly rotated the actual galaxies from the survey, and
added the shear field from the simulation.  The random rotation erases
all two point information. 
We generated a spin-2 Gaussian random fields consisting of a pure E
mode, and sampled that field at the galaxy positions.  The mock
catalog consists of the original galaxy rotated by a random angle, and
with its ellipticity divided by 2 to present a more significant
challenge to the condition number.  To this we added the Gaussian signal.
Our multiscale procedure converged to machine precision after 8
iterations in double precision.

\begin{figure}
\vskip 3.3 truein
\includegraphics{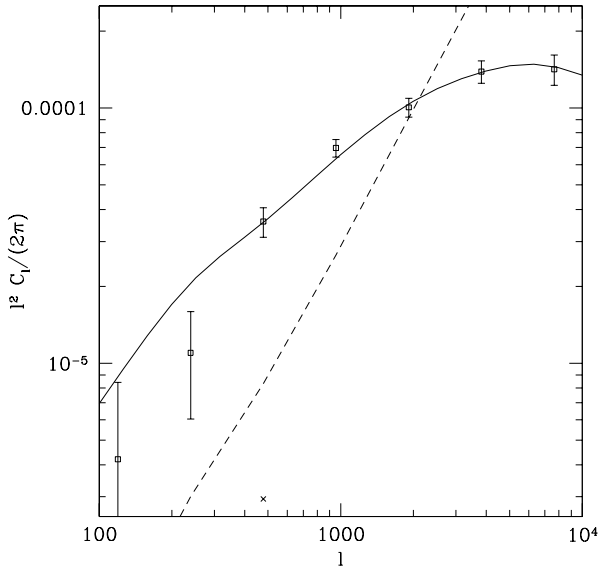}
\caption{
Angular power spectrum recovered from a mock lensing field.
The solid line shows the input power spectrum, the dashed line
is the quadratic estimator of the noise.  The boxes with error bars
indicate the recovered power from the noisy catalogs, with the crosses
indicating the recovered B mode.  The simulations contained no B mode.
}
\label{fig:clsim}
\end{figure}

Figure \ref{fig:clsim} shows the angular power spectrum recovered from a
simulated galaxy survey.
We see that the power spectrum of the noise crosses from small at
large scales to dominant for $l>1000$, so a pure noise weighted
correlation analysis is non-optimal at large angular scales.

\section{Conclusion}

We have presented the framework to rapidly and optimally compute power
spectra.  Its computational cost is $O(N \log N)$, and involves a
stochastic evaluation of traces.  The latter converges to the exact
result in $N$ iterations, yielding the exact solution in $O(N^2 \log
N)$ operations.  In practice, many fewer iterations provide the answer
to sufficient accuracy.  The stochastic evaluation is also trivially
run on a cluster of cheap computers, which are often readily
available.

We have presented mathematical estimates of convergence, which is good
in most practical applications.  For the estimation of weak lensing
power, only a few iterations are required to converge to machine
precision.

For the upcoming lensing and CMB surveys such as the
Canada-France-Hawaii-Telescope-Legacy-Survey, this rapid algorithm
will allow an optimal analysis of the data in a tractable amount of
computational effort.

I would like to thank Yannick Mellier and Ludovic van Waerbeke for
providing the Virmos galaxy data sets.

\bibliography{penbib}

\begin{thebibliography}{}

\bibitem[\protect\citeauthoryear{{Bond}}{{Bond}}{1995}]{1995PhRvL..74.4369B}
{Bond} J.~R.,  1995, Physical Review Letters, 74, 4369

\bibitem[\protect\citeauthoryear{{Bunn} \& {White}}{{Bunn} \&
  {White}}{1997}]{1997ApJ...480....6B}
{Bunn} E.~F.,  {White} M.,  1997, \apj, 480, 6

\bibitem[\protect\citeauthoryear{{Feldman}, {Kaiser} \& {Peacock}}{{Feldman}
  et~al.}{1994}]{1994ApJ...426...23F}
{Feldman} H.~A.,  {Kaiser} N.,    {Peacock} J.~A.,  1994, \apj, 426, 23

\bibitem[\protect\citeauthoryear{{Hockney} \& {Eastwood}}{{Hockney} \&
  {Eastwood}}{1988}]{1988csup.book.....H}
{Hockney} R.~W.,  {Eastwood} J.~W.,  1988, {Computer simulation using
  particles}.
Bristol: Hilger, 1988

\bibitem[\protect\citeauthoryear{{Hu} \& {White}}{{Hu} \&
  {White}}{2001}]{2001ApJ...554...67H}
{Hu} W.,  {White} M.,  2001, \apj, 554, 67

\bibitem[\protect\citeauthoryear{{Oh}, {Spergel} \& {Hinshaw}}{{Oh}
  et~al.}{1999}]{1999ApJ...510..551O}
{Oh} S.~P.,  {Spergel} D.~N.,    {Hinshaw} G.,  1999, \apj, 510, 551

\bibitem[\protect\citeauthoryear{{Padmanabhan}, {Seljak} \&
  {Pen}}{{Padmanabhan} et~al.}{2002}]{2002astro.ph.10478P}
{Padmanabhan} N.,  {Seljak} U.,    {Pen} U.~L.,  2002, in eprint
  arXiv:astro-ph/0210478 {Mining Weak Lensing Surveys}.
pp 10478--+

\bibitem[\protect\citeauthoryear{{Pen}, {Van Waerbeke} \& {Mellier}}{{Pen}
  et~al.}{2002}]{2002ApJ...567...31P}
{Pen} U.,  {Van Waerbeke} L.,    {Mellier} Y.,  2002, \apj, 567, 31

\bibitem[\protect\citeauthoryear{{Pen}, {Zhang}, {Waerbeke}, {Mellier}, {Zhang}
  \& {Dubinski}}{{Pen} et~al.}{2003}]{2003astro.ph..2031P}
{Pen} U.,  {Zhang} T.,  {Waerbeke} L.~v.,  {Mellier} Y.,  {Zhang} P.,
  {Dubinski} J.,  2003, in eprint arXiv:astro-ph/0302031 {Detection of dark
  matter Skewness in the VIRMOS-DESCART survey: Implications for
  $\backslash$Omega\_0}.
pp 2031--+

\bibitem[\protect\citeauthoryear{{Seljak}}{{Seljak}}{1998}]{1998ApJ...506...64%
S}
{Seljak} U.,  1998, \apj, 506, 64

\bibitem[\protect\citeauthoryear{{Szapudi}, {Prunet}, {Pogosyan}, {Szalay} \&
  {Bond}}{{Szapudi} et~al.}{2001}]{2001ApJ...548L.115S}
{Szapudi} I.,  {Prunet} S.,  {Pogosyan} D.,  {Szalay} A.~S.,    {Bond} J.~R.,
  2001, \apjl, 548, L115

\end{thebibliography}
\bibliographystyle{mn2e}
\appendix

\bsp
\label{lastpage}

\end{document}